\documentclass[onecolumn,superscriptaddress,prl]{revtex4}

\usepackage{epsfig}
\usepackage{amsfonts}
\usepackage{graphicx}% Include figure files
\usepackage{dcolumn}% Align table columns on decimal point
\usepackage{bm}% bold math
\usepackage{lettrine}

\begin{document}

\title{Understanding the nature of "superhard graphite"}

\author{Salah Eddine Boulfelfel}
\email{sboulfelfel@notes.cc.sunysb.edu}
\affiliation{Stony Brook University, Department of Geosciences, NY 11794-2100, USA}
\author{Artem R. Oganov}
%\email{aoganov@notes.cc.sunysb.edu}
\affiliation{Stony Brook University, Department of Geosciences, NY 11794-2100, USA}
\affiliation{Department of Geology, Moscow State University, 119899 Moscow, Russia}
\author{Stefano Leoni}
\affiliation{Technische Universit\"at Dresden, Institut f\"ur Physikalische Chemie, 01062 Dresden, Germany}
%\email{stefano.leoni@chemie.tu-dresden.de}

\begin{abstract} 
Numerous experiments showed that on cold compression graphite transforms into a new superhard and transparent allotrope.
Several structures with different topologies have been proposed for this phase.
While experimental data are consistent with these models, the only way to solve this puzzle is to find which structure is kinetically easiest to form.
Using state-of-the-art molecular-dynamics transition path sampling simulations,
we investigate kinetic pathways of the pressure-induced transformation of graphite to various superhard candidate structures.
Unlike hitherto applied methods for elucidating nature of superhard graphite, transition path sampling realistically models nucleation events necessary for physically
meaningful transformation kinetics.
We demonstrate that nucleation mechanism and kinetics lead to $M$-carbon as the final product.
$W$-carbon, initially competitor to $M$-carbon, is ruled out by phase growth.
Bct-C$_4$ structure is not expected to be produced by cold compression due to less probable nucleation and higher barrier of formation.
\end{abstract}

\maketitle

\lettrine[lines=1,slope=-4pt,nindent=-4pt]{C}omputational materials design is a central grand challenge of modern science.
Recent progress in the development of theoretical methodologies for structure prediction, especially evolutionary algorithms~\cite{AROBook},
has led to important discoveries confirmed by subsequent experiments~\cite{AROB,ARONa}.
In addition to thermodynamically stable solid phases, metastable modifications are worth exploring
both computationally and experimentally because of their sometimes superior physical properties.
However, among the infinite number of all possible metastable phases for each compound, only a handful are synthesizable for kinetic reasons.
A metastable phase is synthesizable if it has the highest kinetic likelihood of formation at given conditions.

The recently revived debate~\cite{MC,BCT4,NEBC,WC,ZC1,ZC2} on the nature of the metastable product of cold compression of graphite~\cite{G1,G2,H5,H7} represents a special case of interest.
The transformation of graphite to the thermodynamically stable cubic diamond occurs at high pressures and temperatures ($>$ 5 GPa and 1200-2800 K) and in the presence of catalysts.
At ambient temperature, however, a new phase is formed above 15-19 GPa~\cite{G1,G2,H5,H7} and the transition is accompanied by clear changes in the physical properties of carbon~\cite{MC,WC}.

Experiments have reported an increase in the electrical resistivity~\cite{H2,H3,H4,H5,H6,H1} and optical transmittance~\cite{H3,H5}.
Raman spectra measurements up to 14 GPa indicate that the frequencies of the $E_{2g}^{(1)}$ rigid-layer shear mode at 44 cm$^{-1}$ and the $E_{2g}^{(2)}$ in-plane mode at 1579 cm$^{-1}$
increase sublinearly under pressure with initial pressure coefficients of 4.8 and 4.7 cm$^{-1}$/GPa, respectively~\cite{H1}.
Changes indicating the formation of a new phase have been also observed in the x-ray diffraction pattern~\cite{H2,H4,H6,H7} and in the near $K$-edge spectroscopy~\cite{H7}.
More remarkable is the increase in hardness, as evidenced by the ability of the new phase to indent diamond anvils~\cite{H7}.

The difficulty to resolve the crystal structure of this mysterious superhard phase, which we call "superhard graphite", from experiments has stimulated theoretical efforts~\cite{MC,WC,NEBC,BCT4}.
Hexagonal and cubic diamond have been ruled out because of the absence of their characteristic Raman bands at 1330 cm$^{-1}$ \cite{RAMN}.
Amorphization was dismissed because of the persistence of Bragg peaks traceable to the original graphite pattern \cite{H7,MC}.

Theoretically, candidate structures with distinct topologies featuring different patterns, combining odd (5 and 7) and even (4, 6, and 8) rings, have been proposed.
A low-enthalpy monoclinic structure called {\it M}-carbon (space group $C2/m$, Z=16) was discovered using evolutionary algorithm USPEX~\cite{aroglass} and was later identified with "superhard graphite"~\cite{MC}.
All carbon atoms in this structure are four-coordinate and form 5-membered and 7-membered rings when the structure is projected on (010).
Later, a body-centered tetragonal structure called bct-C$_4$ (space group $I/4mmm$, Z=8)~\cite{BCT4,NEBC,SMC4} was proposed as another candidate.
The four-coordinate carbon atoms form 4-membered and 8-membered rings on the (001) projection.
Shortly afterwards, an orthorhombic candidate structure called {\it W}-carbon (space group {\it Pnma}, Z=16) was proposed~\cite{WC}.
This structure is very similar to that of {\it M}-carbon, featuring 5-membered and 7-membered rings when the structure is projected on (010).
Very recently, an orthorhombic structure, called {\it o}C16-II ($Cmmm$, Z=16), combining even-membered rings (4, 6, and 8) was predicted using particle-swarm optimization method~\cite{ZC1}
and metadynamics simulations~\cite{ZC2}.
Using the latter method, it was demonstrated that a large variety of superhard structures can be expected ({\it e.g.} {\it o}C16-I, {\it m}C12, and {\it m}C32)~\cite{ZC2}.
%%All of the candidate structures proposed so far, and some of the new ones, were found using the newly developed evolutionary metadynamics method~\cite{QZPRB,QZMETA}.

The simulated x-ray diffraction patterns of the predicted structures ({\it M}-carbon, {\it W}-carbon, bct-C$_4$, and {\it o}C16-II)
are in satisfactory agreement with experimental data~\cite{MC,BCT4,WC,ZC2}.
Their respective bulk moduli (and hardness~\cite{hardness}) are 392.6 (82.7), 391.8 (83.1), 393.4 (82.0), and 408.4 (84.4) GPa~\cite{ZC2}.
Band gaps calculated using the {\it GW} method~\cite{gw} are 5.00~\cite{QZ}, 4.39~\cite{WC}, and 3.78 eV~\cite{BCT4}, for {\it M}-carbon, {\it W}-carbon, and bct-C$_4$, respectively.
All candidate structures show no imaginary phonon frequencies and can explain the experimental observations of hardness and transparency of cold compressed graphite.

Physical properties and energetics of these phases have been theoretically well studied~\cite{MC,WC,BCT4,NEBC,prp1,prp2},
but the existing evidence is insufficient to decide which one was obtained in experiments~\cite{G1,G2,H5,H7}.
In order to clearly identify the product of cold compression of graphite, we need to explore graphite transformation kinetics and compute energy barriers for transition regimes characterized by phase coexistence. 

Investigations of transition routes to superhard graphite~\cite{NEBC,WC} using nudged elastic band method provided inconsistent results, especially inferring that
graphite to cubic diamond has lower transition barrier than transformations to other three possible structures~\cite{WC}.
In addition, Ref.~\cite{WC} reports a different sequence of energy barriers (cubic diamond $<$ {\it M}-carbon $<$ bct-C$_4$) than Ref.~\cite{NEBC} (bct-C$_4$ $<$ cubic diamond $<$ {\it M}-carbon),
despite the use of the same computational method.
These conflicting results highlight the intrinsic complexity of a reliable elucidation of transition mechanisms and associated barriers.

In a complex system, like a three dimensional solid, the lowest energy state of a system is not always the most relevant \cite{MOK}.
The viability of any candidate structure as a metastable product is strictly connected to the corresponding transformation pathway and the associated transition barriers
or dynamical bottlenecks \cite{AROBook,salah1,SL1}.
Transition barriers can be computed and bottlenecks identified with (stationary) saddle points representing transition states on the potential energy surface \cite{FSBook}.
In complex systems with very rugged energy landscapes, saddle points cease to be characteristic points of the free energy barrier \cite{TPSB1}.
The latter barrier encompasses a large set of configurations and the associated relevant degrees of freedom are very difficult to anticipate \cite{FSBook,TPSB1,AROBook}.
To overcome this problem, we can use transition path sampling, a method based on collecting ensembles of dynamical trajectories \cite{AROBook,salah1,SL1,FSBook,TPSB1}.

\section{Results}

In this work, we elucidate the nature of the metastable product of cold compression of graphite and compute the associated energy barrier from true dynamical pathways.
The mechanistic details of this pressure-induced phase transformation are determined at the atomic level with full account for relevant events of nucleation.
To achieve a reliable and detailed mechanistic picture of the transition we employ a strategy that combines isothermal-isobaric ({\it NpT}) molecular dynamics with
transition path sampling (TPS)~\cite{TPS1,TPS2,TPS3} (see {\it Methods}).
This methodology has proven very effective in the simulation of activated processes~\cite{AROBook,FSBook,SL3} with phase coexistence and growth \cite{salah1,SL1,salah3}.

{\bf {\em How TPS works}}
Transition path sampling is a generalization of Monte Carlo procedures in the space of dynamical trajectories connecting two states (graphite and metastable superhard phase in this case)
separated by a high barrier in a rough energy landscape \cite{TPS1,TPS2,TPS3}.
Therein, the relevance of a transition pathway is biased by path probability.
The overall simulation approach is iterative and develops from an initial trajectory.
By analogy with conventional MC simulations, the first step of TPS consists of equilibrating an initial pathway and gradually converging the trajectory regime to more probable regions.
Accordingly, the first trajectory does not need to be a probable one or include mechanistic details of the real transition regime.

In a situation characterized by the possibility of many theoretically viable candidate structures, TPS is a very helpful simulation strategy because it allows starting from a regime of very
low probability or even a model constructed from a mapping between the limiting phases of the transition of interest.
The convergence is ensured by Monte Carlo sampling of the space of trajectories.
In the case of cold compression of graphite, we started from a pathway connecting graphite to cubic diamond generated by propagation of configurations obtained
from a modeling approach based on transforming periodic nodal surfaces calculated from a short Fourier summation for each structure~\cite{pns1,pns2} (see {\it Methods}).

{\bf {\em Finding the real transition route}}
The importance of nucleation events during the compression of graphite was pointed out for the transition to diamond at high pressures and temperatures~\cite{parri}.
In the following, we investigate the kinetics of graphite transformation at ambient temperature due to intrinsic differences between nucleation and growth patterns of predicted
superhard structures. 

The first set of TPS runs was started from trajectories connecting graphite to cubic diamond.
Since diamond (cubic and hexagonal) was ruled out by experiments~\cite{RAMN} and due to high activation barrier separating diamond from graphite, starting with this pathway
will not only optimize the search in the path ensemble but will help converging to a more probable end point structure.
The appearance of a new different phase instead of diamond is allowed by setting a flexible order parameter that can accommodate four-coordinate structures (see {\it Methods}). 

The set of transition pathways harvested in the course of TPS iterations shows a quick departure of the trajectory regime from the collective motion inherited from the geometrical modeling.
The graphite-diamond trajectories underwent three drastic shifts in the transition regime.

\begin{figure*}[!]
\epsfig{file=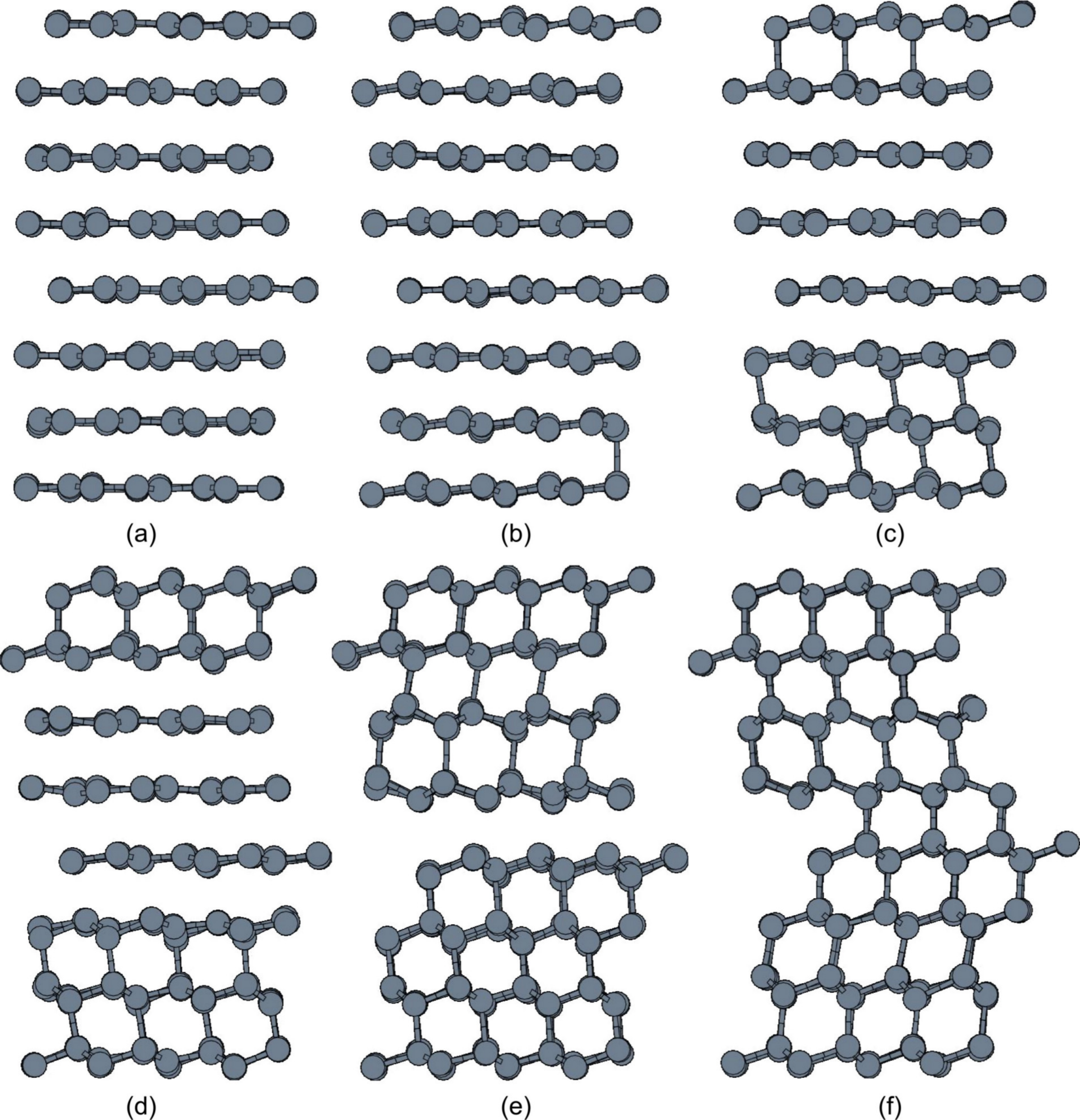,width=1\textwidth,keepaspectratio}
\caption{Snapshots from a dynamical trajectory collected from transition path sampling connecting (a) graphite to (f) a polytype intermediate between
cubic and hexagonal diamond. The buckling of graphite layers in initiated by the formation of C$-$C bonds along [001]$_{graphite}$ (b)-(c).
Domains of cubic diamond are formed with different orientation (d)-(e). The interface between the latter domains defines a region of hexagonal diamond (e)-(f)}
\label{fig:figure1}
\end{figure*}

First, after 15-20 iterations, the end point of the initial trajectory, graphite to cubic diamond, is replaced by a polytype intermediate between hexagonal and cubic diamond
(Fig. \ref{fig:figure1}).
The appearance of this polytype is due to opposite sliding directions of graphene layers giving rise to layers of hexagonal diamond.
This evolution of the transition regime is similar to the results of a molecular dynamics investigation of the pressure-induced transformation path of graphite of diamond
at high temperatures \cite{MDC}.

\begin{figure*}[!]
\epsfig{file=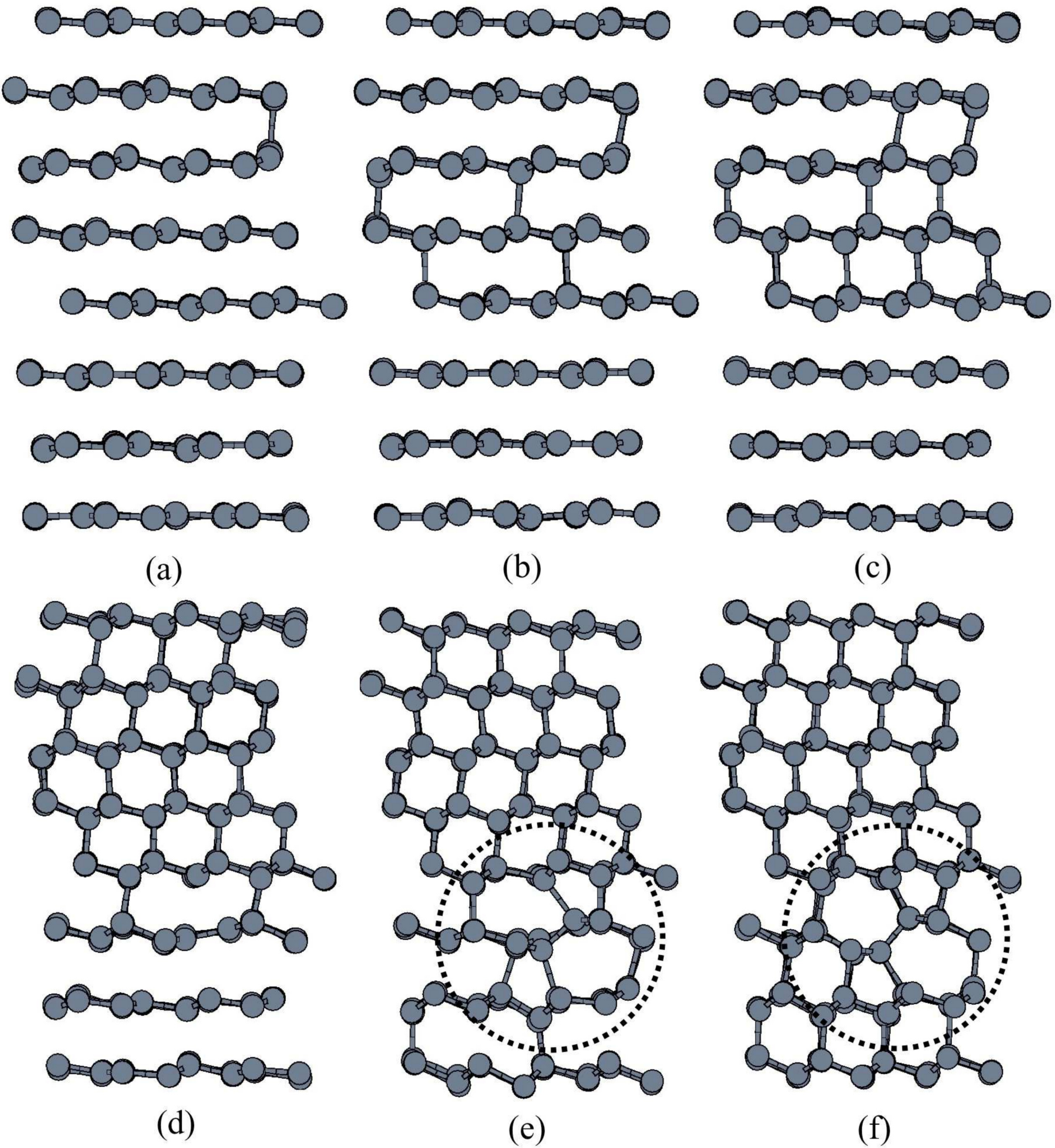,width=1\textwidth,keepaspectratio}
\caption{Snapshots from a representative trajectory illustrating the evolution of the (a) graphite to (f) cubic diamond transition regime.
The mobility of graphitic layers during the reconstruction creates (d)-(f) an inset of 5- and 7-membered rings (dotted circle)
within a 6-membered rings network.
This inset interfaces well with cubic diamond and represents the seed of the metastable phase resulting from the cold compression of graphite.}
\label{fig:figure2}
\end{figure*}

After 30-40 iterations, an inset of new structural patterns made of 5-membered and 7-membered rings appeared within 6-membered rings network (Fig. \ref{fig:figure2}, (d)-(f)).
These new patterns are still made of four-coordinate carbon atoms and interface very well with cubic diamond.
They are characteristic of two structures suggested as metastable product for the cold compression of graphite, {\it M}-carbon and {\it W}-carbon.
The small size of the inset, however, does not yet allow to discriminate between these two models (Fig. \ref{fig:figure2}, (f), dotted circle).

\begin{figure*}[!]
\epsfig{file=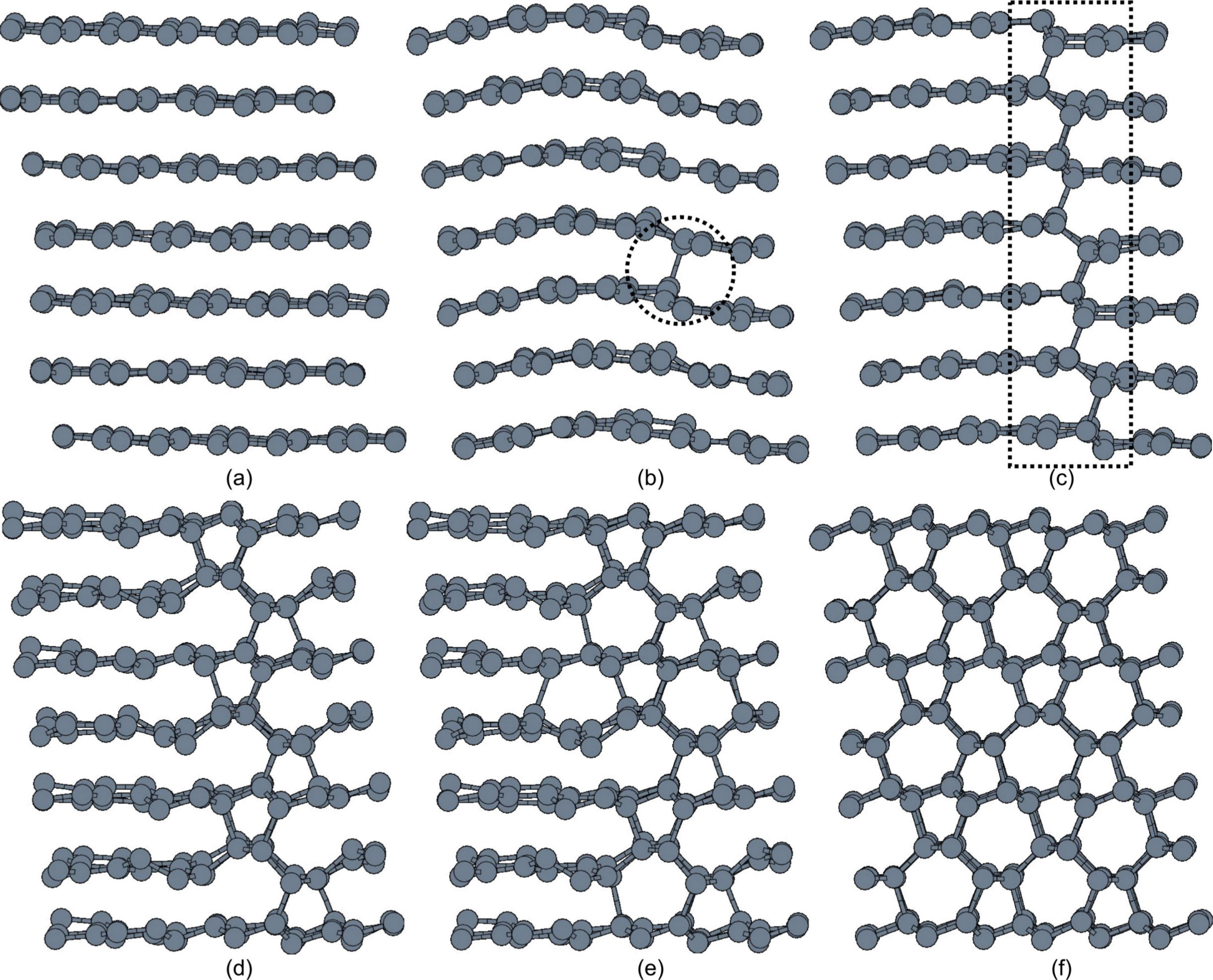,width=1\textwidth,keepaspectratio}
\caption{Snapshots of a representative trajectory of the stable regime corresponding to the cold compression of graphite.
A single event of (a) bond formation (dotted circle) between graphitic layers triggers (b) a series of bond formation along [001]$_{graphite}$ in a zigzag fashion (dotted rectangle).
The latter contacts facilitate the formation of (c)-(d) 5-membered rings causing the corrugation of graphitic layers and inducing the formation
of (d)-(e) 7-membered rings.}
\label{fig:figure3}
\end{figure*}

Upon further sampling of the pathway ensemble, the inset of odd-membered rings grew larger and converted the remaining cubic diamond regions into a structure identical to that of {\it M}-carbon.
The new transition regime connects graphite to {\it M}-carbon (Fig. \ref{fig:figure3}).
Therein, the onset of the transformation is marked by the formation of interlayer single C$-$C bond (Fig. \ref{fig:figure3}, (a)).
The formation of the nucleus triggers a series of bond formation events along [001]$_{graphite}$ in a zigzag fashion (Fig. \ref{fig:figure3}, (b)).
The breaking of $\pi$-interactions within graphitic layers by the emergence of the latter zigzag chain induces layer corrugation that facilitates the formation of 5-membered rings
(Fig. \ref{fig:figure3}, (c)-(d)).
The latter rings pair up, and additional 7-membered rings are formed (Fig. \ref{fig:figure3}, (e)).

The degrees of freedom induced by the mobility of graphene layers under pressure theoretically allow to consider different ways of graphene layers buckling.
The survival of an initial regime or the appearance of a new one is connected to the way of nucleating the high pressure modification and the corresponding free energy barrier.
TPS is biased to efficiently converge to the relevant set of the path ensemble which corresponds to the real transition pathway regime \cite{TPS1,TPS2,TPS3,SL1,SL3} and bypasses
less-favored regions of the energy landscape corresponding to other mechanisms, graphite-bct-C$_4$ and graphite-{\it W}-carbon in this case.
To illustrate this feature of TPS, we performed separate simulation runs to evaluate the energy barrier of graphite to bct-C$_4$ and to {\it W}-carbon transition routes
and clarify the associated reconstruction mechanisms in comparison with the {\it M}-carbon route.

{\bf {\em Viability of graphite to W-carbon pathway}}
Despite its structural and energetic closeness to the {\it M}-carbon phase \cite{WC}, the arrangement of odd-membered rings corresponding to the orthorhombic candidate
structure, {\it W}-carbon, only appears in 20 $\%$ of trajectories collected during TPS simulations performed starting from graphite to diamond trajectories, over a pressure range from 15$-$20 GPa.

\begin{figure*}[!]
\epsfig{file=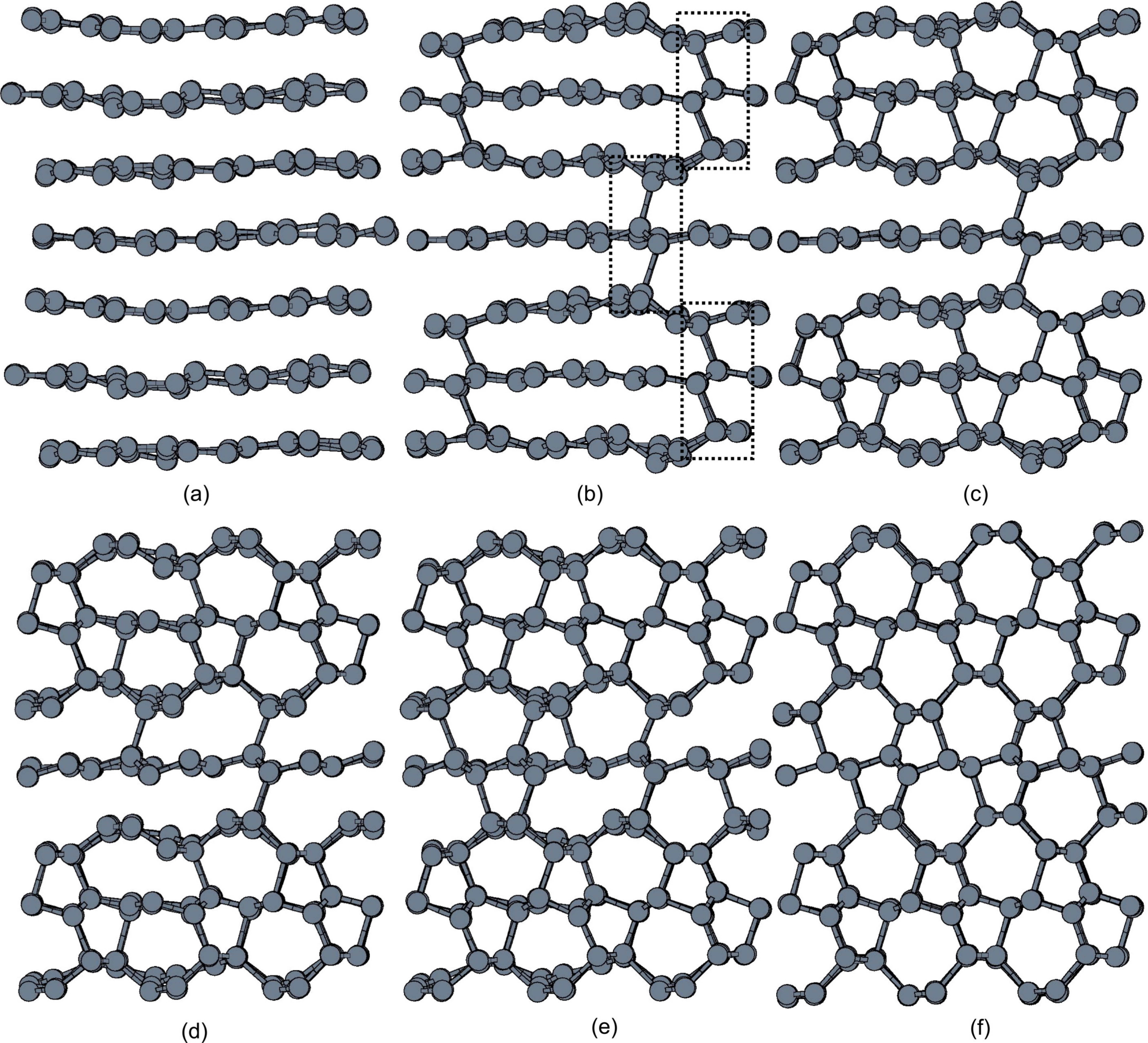,width=1\textwidth,keepaspectratio}
\caption{Snapshots taken from a representative graphite to {\it W}-carbon transformation pathway.
The buckling of graphite is initiated by (a)-(b) formation C$-$C contacts along [001]$_{graphite}$ in the form of finite size zigzag chains.
Each zigzag chain facilitates the corrugation of graphitic layers inducing the formation of (c)-(e) 5- and 7-membered rings transforming graphite into (f) {\it W}-carbon.}
\label{fig:figure4}
\end{figure*}

In order to investigate differences in terms of transition mechanism and kinetics, we modeled a distinct set of TPS runs.
The starting trajectory was taken from the previous runs that converged to graphite to {\it M}-carbon.
Although nucleation events triggering the transition along the graphite to {\it W}-carbon path (Fig.~\ref{fig:figure4}, (a)-(b)) are very similar to that of graphite to {\it M}-carbon
(Fig.~\ref{fig:figure3}, (a)-(d)), subsequent growth of {\it W}-carbon slightly differs from that of {\it M}-carbon.
The formation of a C$-$C bond bridging two graphitic layers, in the case of {\it W}-carbon, induced the appearance of other bonds along [001]$_{graphite}$ as shown in Fig.~\ref{fig:figure4}.
Unlike in the transition mechanism to {\it M}-carbon, the zigzag chain propagates in a segmented fashion (Fig.~\ref{fig:figure4}(b)).
The key step in the evolution of graphite toward {\it M}-carbon or {\it W}-carbon is the evolution of a single nucleation event, formation of an inter-layer C-C bond,
either as infinite zigzag (Fig.~\ref{fig:figure3}(c)) or segmented chain (Fig.~\ref{fig:figure4}(b)).
The breaking of the chain implies different progress of phase growth and results in a higher transition barrier (see below for transition kinetics).
The average length of graphite to {\it W}-carbon transformation is 9 ps while graphite to {\it M}-carbon is 8 ps or less.
This reflects slower transition rate for {\it W}-carbon.

The effect of strain caused by the nucleus on the graphite lattice upon the direct conversion of graphite into diamond has been pointed out to be responsible of the formation of metastable
hexagonal modification rather than cubic in agreement with experimental observations~\cite{parri}.
The nucleus formation along the graphite to {\it W}-carbon transition induced larger distortions of the graphite structure at the interface if compared to the path to {\it M}-carbon.
This explains the more frequent appearance of the latter structure during the sampling. 

{\bf {\em Viability of graphite to bct-C$_4$ pathway}}
As demonstrated above, the graphite to bct-C$_4$ path is a less favored route.
The set of sampling runs of such a trajectory ensemble quickly shift the transition regime toward the more favorable one, graphite-{\it M}-carbon.
To prevent the disappearance of the bct-C$_4$ structure, additional constraints are introduced on the order parameter to keep the transition regime as a pure
graphite to bct-C$_4$ transformation (see {\it Methods}).

\begin{figure*}[!]
\epsfig{file=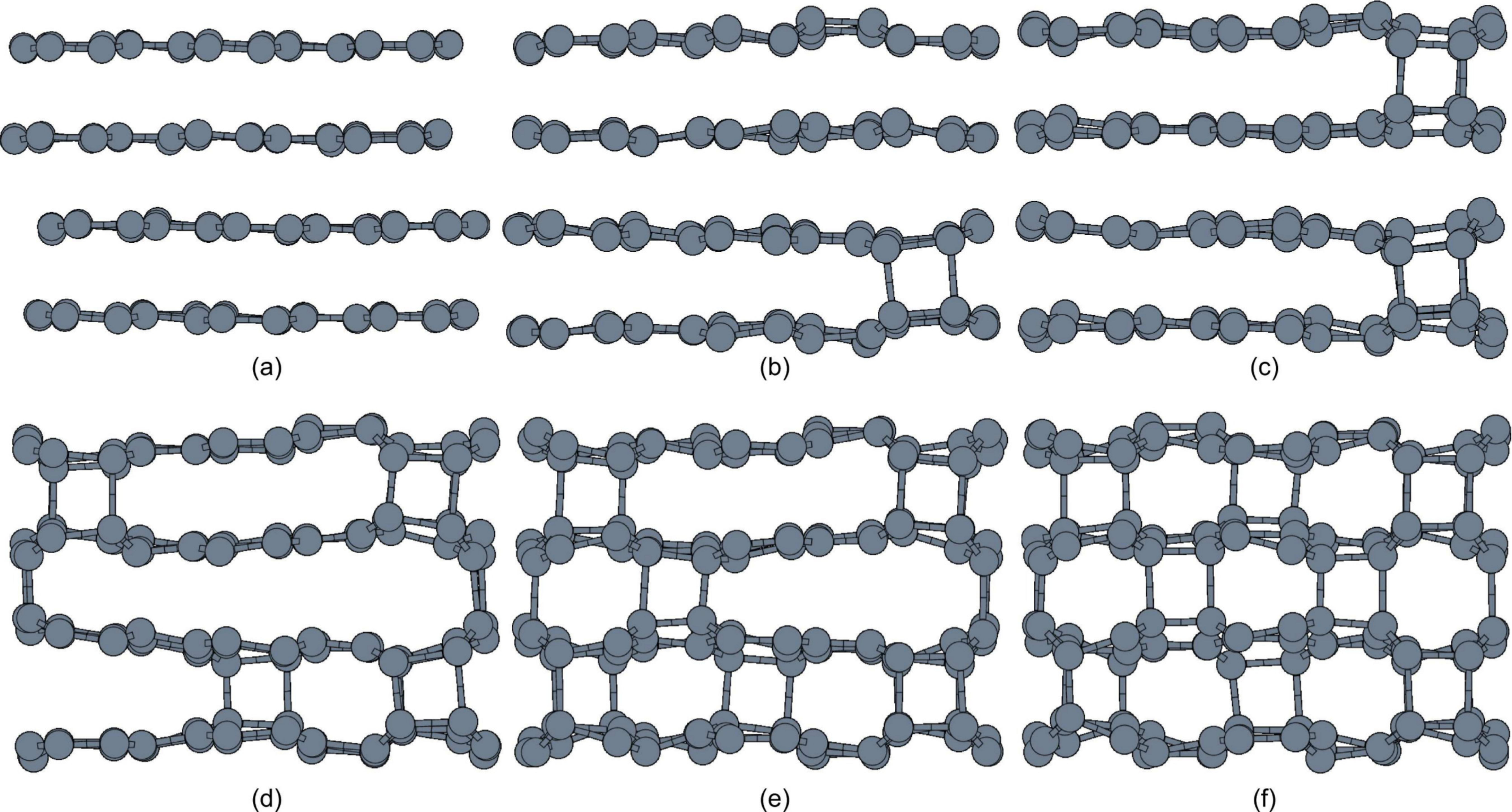,width=1\textwidth,keepaspectratio}
\caption{Snapshots taken from a representative graphite to bct-C$_4$ transformation pathway.
The transition proceeds via nucleation of C$_4$ square units. Further growth is achieved by corrugation of graphitic layers to form more C$_4$ units and reconstruct
the graphite into bct-C$_4$.}
\label{fig:figure5}
\end{figure*}

A representative trajectory of the latter path is depicted in Fig. \ref{fig:figure5}. Unlike the graphite-{\it M}-carbon transition mechanism,
the onset of the reconstruction is marked by the formation of two C$-$C bonds defining a C$_4$ square unit (Fig. \ref{fig:figure5}, (a)-(b)).
The perturbation induced by the C$_4$ nucleus helps propagating this motif and the bct-C$_4$ phase grows until complete reconstruction of graphite into the
body-centered tetragonal structure.

\begin{figure*}[!]
\epsfig{file=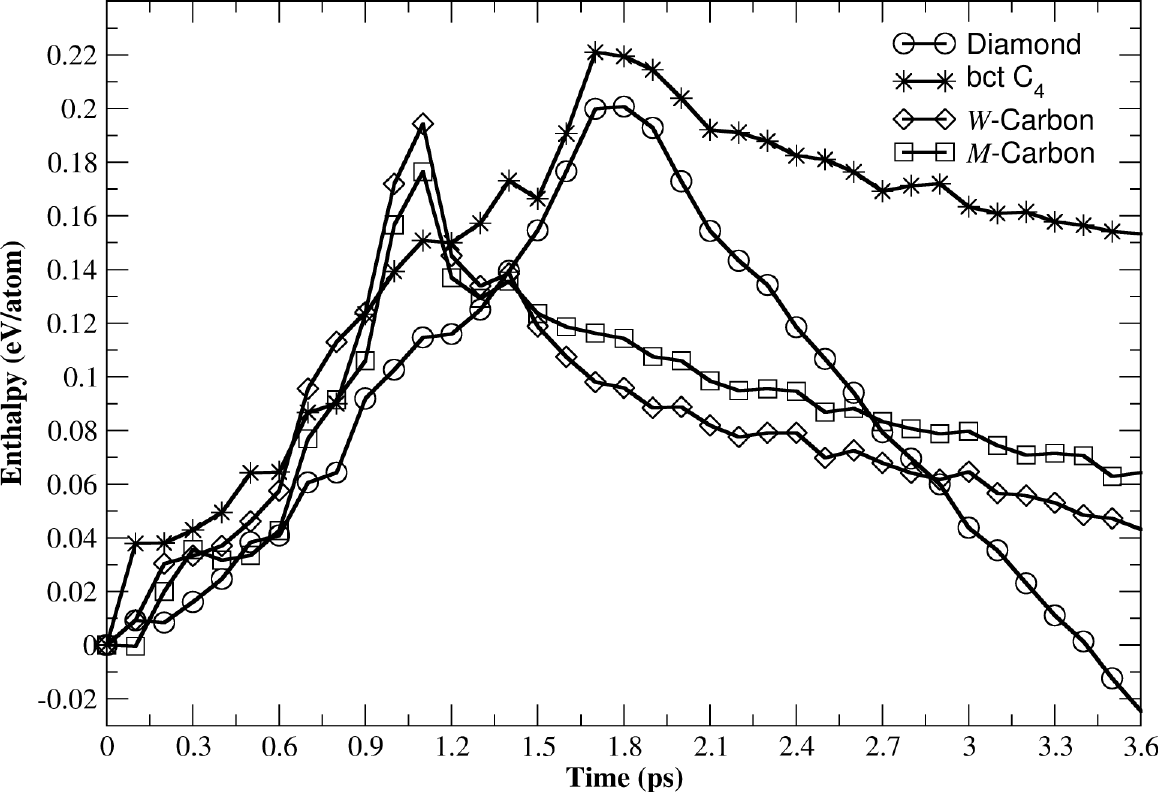,width=1\textwidth,keepaspectratio}
\caption{Enthalpy variation of different simulated transformations of graphite under pressure (15 GPa) and ambient temperature.
The graphite to {\it M}-carbon transformation route (square line) indicates a lower energy barrier than the transition to $W$-carbon
(diamond line).
The possibility of graphite transformation into bct-C$_4$ structure (star line) on cold compression is ruled out because of higher barrier than the
graphite to cubic diamond transition (circle line).}
\label{fig:figure6}
\end{figure*}

{\bf {\em Kinetics of graphite cold compression}}
In order to shed light on the kinetics of the transformation of graphite under high-pressure and ambient temperature, enthalpy variation of different simulated
transition regimes starting from graphite have been computed and depicted in Fig. \ref{fig:figure6}.
The graphite to bct-C$_4$ transition barrier (Fig.~\ref{fig:figure6}, star line) scores as the highest with 221 meV/atom.
The graphite to cubic diamond transformation (Fig. \ref{fig:figure6}, circle line) has a barrier of 200 meV/atom which completely rules out the latter transition route
for graphite cold compression.
Despite the similarities between transformation routes to {\it W}-carbon (Fig. \ref{fig:figure6}, diamond line) and {\it M}-carbon (Fig. \ref{fig:figure6}, square line),
the latter is favored with a barrier equal to 176 meV/atom, lower than 194 meV/atom for the former transition.

\begin{figure*}[!]
\epsfig{file=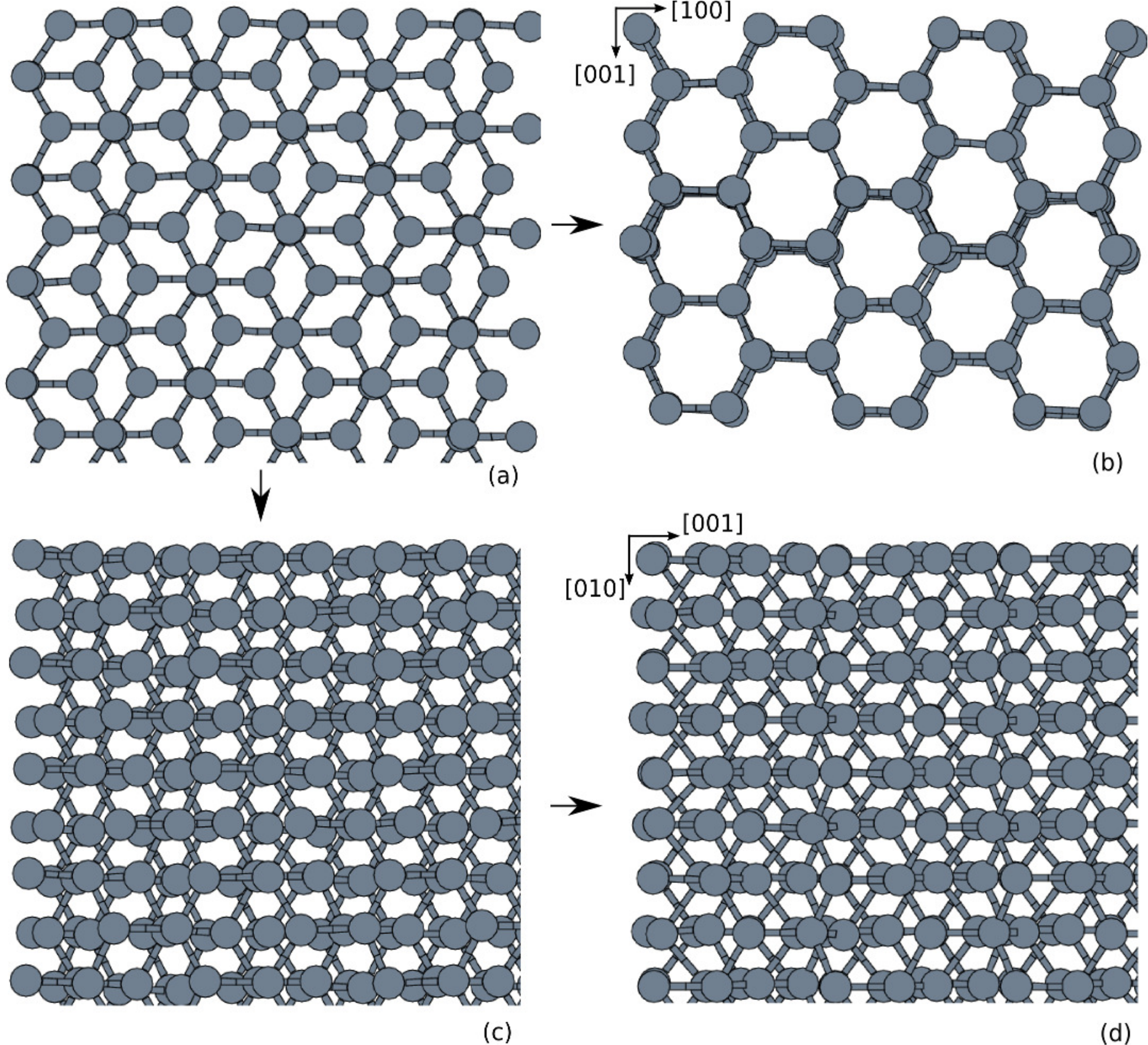,width=1\textwidth,keepaspectratio}
\caption{Top view of graphitic layers sliding fashions parallel to (001) plane.
The (a) graphite to (b) bct-C$_4$ transition requires an eclipsed arrangement of next-neighboring layer along $[001]$ and changes the layers stacking sequence from
...$AB$.. into ...$AA$... in order to trigger layers buckling (only 3 graphitic layers are shown for better clarity).
The (a) graphite to (d) {\it M}-carbon transformation implies small atomic displacements to induce (c) the onset of graphitic layers buckling.}
\label{fig:figure7}
\end{figure*}

{\bf {\em The role of graphite layers sliding}}
In the nucleation process, the buckling of graphitic layers is realized upon their corrugation followed by formation of new C-C bonds
(Fig.~\ref{fig:figure1}-(b), Fig.~\ref{fig:figure2}-(a), Fig.~\ref{fig:figure3}-(b), Fig.~\ref{fig:figure4}-(a), and Fig.~\ref{fig:figure5}-(b)).
This is facilitated by layer sliding in graphite that brings carbon atoms into an appropriate stacking sequence for C-C bonds formation along $[001]$.
Just prior to nucleus formation, the applied pressure causes graphite layers to slide in a fashion that determines the nature of the high pressure modification to be formed.
In Fig.~\ref{fig:figure6}, the variation of enthalpy indicates a different evolution for graphite even before the formation of any nucleus for all investigated transformations
(graphite to cubic diamond, to bct-C$_4$, to {\it W}-carbon, and to {\it M}-carbon).
For $t=0-600$ fs, graphite to bct-C$_4$ shows a higher enthalpy kink if compared to the two other candidates, {\it W}- and {\it M}-carbon.
This slight difference in enthalpy (40 meV/atom for bct-C$_4$ and 30 meV/atom for {\it W}- and {\it M}-carbon) reflects two different layer sliding mechanisms as shown in Fig.~\ref{fig:figure7}.
The graphite to bct-C$_4$ transition requires an eclipsed arrangement of next-neighboring layer along $[001]$ and changes the layers stacking sequence
from ...{\it AB}... into ...{\it AA}... (Fig.~\ref{fig:figure7}, (a)-(b)) in order to trigger layers buckling.

The transformations from graphite to 5+7 structures on the contrary require a reduced amount of displacements to induce graphitic layer buckling
(Fig.~\ref{fig:figure7}, (a)-(d)).
Therefore, transformation routes leading to {\it W}- and {\it M}-carbon are more favored than the path to bct-C$_4$ in terms of layer sliding mechanism in graphite.
Additionally, pathways to {\it W}- and {\it M}-carbon are less demanding in terms in-layer distortions, necessary to initiate nucleation.

The stacking sequence and atomic displacements in the case of pathways to {\it W}- and {\it M}-carbon imply minimal in-layer distortions in order to initiate nucleation.
The path to {\it M}-carbon is favored over the one to {\it W}-carbon because of different growth processes making the former structure more probable to form.

The effect of simulation box size has been pointed out in large-scale molecular dynamics simulations of graphite to diamond transition \cite{parri}.
If the simulation cell is too small, the derived transition mechanism does not feature in-layer distortions because in-layer stress implies artificial sliding of graphitic layers.
The dimensionality of our simulation box fairly enables the observation of relevant nucleation events and allows to capture distortions around growing nucleus without inducing artificial
layer sliding as shown in Fig~\ref{fig:figure3}-(c) and Fig~\ref{fig:figure4}-(b).
Our results underline the role of distortions around the nucleus and how they determine the kinetics and the overall direction of the transformation.

The very recently proposed candidate {\it o}C16-II deserves special notes.
This orthorhombic structure can be described as an alternation of diamond and bct-C$_4$ slabs along [100] (Fig. $3$ in Supplementary Information).
The existence of 4-membered rings suggests mechanistic features of buckling in a fashion similar to graphite-bct-C$_4$ transformation.
To reconstruct graphite into a structure with a topology of rings fused as 4 $+$ 6 $+$ 8 pattern, atomic displacements in terms of layers sliding are expected to be larger than in the
mechanism leading to {\it M}- and {\it W}-carbon (Fig. $3$ in Supplementary Information).  
On the experimental side, Zhao {\it et al.}~\cite{ZC1} demonstrated an accordance of the calculated x-ray diffraction pattern of {\it o}C16-II
with the experimental XRD data collected for cold compression of carbon nanotube bundles. The inconsistency with experimental data on "superhard graphite" has been pointed out~\cite{ZC1}.
In our TPS simulations, the initial graphite to diamond pathway evolved into the graphite to {\it M}-carbon pathway, bypassing the {\it o}C16-II structure, in spite of its structural closeness to diamond.

\section{Discussion}

Unlike synthesis from the gas phase, the high-pressure synthetic route of superhard graphite depends on the particular nucleation history which would favor
one pattern over another at the stage of phase growth within graphite.
The viability of transformation routes is not determined by the overall stability of a particular candidate structure as an only parameter, but rather by activation energy of formation.
Local events of C-C bond breaking and formation control the kinetics of the transition.
In this work, an atomistic and kinetic picture of cold compression of graphite is presented using a methodology based on molecular-dynamics simulations
combined with transition path sampling.
It illustrates how transition path sampling can be used to assess synthesizability of metastable phases by optimizing transformation pathways.
Unlike static models, we provide an unbiased description of different dynamical routes and evaluation of the associated transition barriers.
The final product of this first-order transformation is unequivocally identified as a metastable monoclinic structure, {\it M}-carbon.
Although a competitor, {\it W}-carbon is kinetically less favored and differences in transformation rates can be traced back to different patterns of phase nucleation
and subsequent growth.
We demonstrate that large distortions caused by the nucleation mechanism of {\it W}-carbon make it less likely to form than {\it M}-carbon.
Another earlier proposed structures, bct-C$_4$, is ruled out because it did not appear during trajectory sampling under cold-compression conditions.

It is worth mentioning that transformation pathways to "superhard graphite" are investigated as homogeneous nucleation processes.
The presence of structural defects in the form of interlayer bonds can facilitate nucleation.
Consequently, heterogeneous nucleation implies lower activation barriers.
However, differences between homogeneous and heterogeneous nucleation do not alter the main features of the transformations revealed in this work.
Phase growth kinetics around lattice distortions and structural defects are expected to be similar to that of homogeneous nucleation model.
Intrinsic differences between different routes to "superhard graphite" are determined by seed growth fashion.
In order to investigate the effect of defects on the kinetics and atomistic mechanism of the transition to "superhard graphite",
transition path sampling molecular dynamics simulations of heterogeneous nucleation will be carried out in the future.
All in all, the investigation conducted in this work offers new insights into the understanding of materials synthesizability under high pressure,
in a situation of competing phases and in absence of better resolved experimental data.

\section{Methods}
\subsection{Molecular Dynamics Simulations}
Born-Oppenheimer molecular dynamics simulations were performed in the {\it NpT} ensemble~\cite{FSBook} at temperature T=300 K and pressures ranging from 15 to 20 GPa.
Constant pressure and temperature were ensured by the Martyna-Tobias-Klein algorithm \cite{npt} allowing for anisotropic shape changes of the simulation box.
Interatomic forces are computed within the framework of density functional tight binding approach~\cite{dftb1,dftb2} as implemented in the CP2K code~\cite{cp2k1}.
For the accuracy of the set of DFTB parameters \cite{dftbC} used, see {\it Supplementary Information}.
The time propagation of the system was performed using the velocity Verlet algorithm~\cite{FSBook}.
To ensure good time-reversibility an integration timestep of 0.2 fs was used.
The simulation box contained 256 carbon atoms.
The starting configurations in all simulations correspond to graphite-2H structure.
The presence of defects and surface effects on the transition mechanism and the nature of the product of the transition is not investigated because the main focus is set on
homogenous nucleation in graphite compressed at ambient temperature.

\subsection{Transition Path Sampling}
The sampling starts from an initial trajectory connecting the limiting phases.
A new trajectory is generated by selecting a configuration from the existing one and slightly modifying the
atomic momenta. The modifications $\delta p$ are applied to randomly chosen pairs of atoms ({\it i,j}) according to:
$\overrightarrow{p}_i^{new} = \overrightarrow{p}_i^{old} + \delta p(\overrightarrow{r}_j - \overrightarrow{r}_i) / |(\overrightarrow{r}_j - \overrightarrow{r}_i)|$ and
$\overrightarrow{p}_j^{new} = \overrightarrow{p}_j^{old} - \delta p(\overrightarrow{r}_j - \overrightarrow{r}_i) / |(\overrightarrow{r}_j - \overrightarrow{r}_i)|$,
keeping both momentum and angular momentum conserved.
The resulting atomic momenta are rescaled by a factor of $\sqrt{E_{kin}^{old} / (\sum_i|\overrightarrow{p}_i^{new}|^2/2m_i)}$, in order to keep the total kinetic energy conserved.
The propagation of the modified configuration in both directions of time (-{\it t},+{\it t}) generates a new trajectory.
Repeating this step provides a set of trajectories and the successful ones are collected and analyzed.
We initialized TPS with trajectories representing four different transformation regimes: graphite$-$cubic diamond, graphite$-$bct-C$_4$, graphite$-${\it M}-carbon,
and graphite$-${\it W}-carbon.
Subsequent equilibration of the path ensemble at 300 K and pressures ranging from 15 to 20 GPa in two simulation runs for each transition regime resulted into 8 independent
TPS simulation runs.
The number of trial shootings for each run exceeded 500 and the acceptance ratio was between 40-60 \%.
This production part of TPS procedure resulted in an ensemble of more than 200 successful pathways, between 8-10 ps long, of which $\sim$70 are independent uncorrelated trajectories.  

\subsection{Order Parameter}
The average coordination number (CN) of carbon atoms is used as an order parameter to distinguish graphite (CN=3) from high pressure modifications made of four-coordinate atoms.
All structures discussed in the manuscript ({\it M}-carbon, {\it W}-carbon, bct-C$_4$, hexagonal and cubic diamond) have an average CN=4 within the first nearest neighbor coordination sphere.
In order to narrow down the path ensemble sampling we used second and third nearest neighbors to distinguish between different four-coordinate structures
Clearly, apart from a quick detection of graphite and different candidate structures (bct-C$_4$, {\it M}-carbon, and {\it W}-carbon), this order parameter is capable
of accommodating not only the three possible metastable phases but allows for starting transition pathways sampling from unlikely trajectories,
like graphite to cubic or hexagonal diamond.
The order parameter does not impose any bias on the evolution of the trajectory sampling.

\subsection{Topological Models}
We model transformation paths by transforming periodic nodal surfaces (PNS)~\cite{AROBook,pns1,pns2}.
The reciprocal space approach implements short Fourier summations to define a family of surfaces according to the formula:
$f(x,y,z)=\sum_{h,k,l}||S_{hkl}||cos(2\pi(hx+ky+lz)-\alpha_{hkl})$.
$S_{hkl}$ is a geometric structure factor and $\alpha_{hkl}$ is the corresponding phase.
The surface corresponding to $f(x,y,z)=0$ is called PNS.
PNS of the transforming phases ({\it A} and {\it B}) are used as starting points for the definition of a geometric model for the transition.
The choice of a common cell with a constant number of atoms and the periodicity of the model ensure commensurability of the two phases.
After transformation of the reflections of the new setting of the cell, the transition can be formulated as a migration from one structure to the other
along a coordinate {\it s} providing weighted linear mixing of the two functions:
$f_{AB}(x,y,z)=sw_Af_A(x,y,z)+(1-s)w_Bf_B(x,y,z),s\in[0,1]$
To derive the transition state, configurations obtained from the geometric model in the range $s\in[0,1]$ are propagated in molecular dynamics simulations at 300 K.
Random velocities are assigned to atoms and the system is left free to evolve toward either {\it A} or {\it B} without imposing any bias.
Reversing the sign of the time coordinate allows reaching the other phase with a finite probability.

\section{acknowledgments}
This work is funded by DARPA (grant N66001-10-1-4037) and NSF (grant EAR-1114313).
SL acknowledges support from DFG for support (SPP 1415).
Calculations were performed at computing facilities of ZIH (TU-Dresden), CFN at Brookhaven National Laboratory (supported by the U.S. Department of Energy, Office of Basic Energy Sciences, under contract No.
DE-AC02-98CH10086), and Supercomputer Center of the Russian Academy of Sciences Skif (Moscow State University).

\section{Author Contributions}
S.E.B., A.R.O., and S.L. designed research, performed simulations, analyzed data, and wrote the manuscript.

\section{Additional information}

{\bf Competing financial interests} The authors declare no competing financial interests.

%% {\bf Supplementary information} accompanies this paper at http://www.nature.com/scientificreports

%% {\bf License:} This work is licensed under a Creative Commons Attribution-NonCommercial-NoDerivative Works 3.0 Unported License.

%% To view a copy of this license, visit http://creativecommons.org/licenses/by-nc-nd/3.0/

%% {\bf How to cite this article:} Boulfelfel, S.E., Oganov, A.R. \& Leoni, S. Understanding the nature of "superhard graphite"

\newpage

\end{document}